# Is Artificial Intelligence the great filter that makes advanced technical civilisations rare in the universe?

**Michael A. Garrett[a] ***

[a]   *Jodrell Bank Centre for Astrophysics, Dept. of Physics & Astronomy, Alan Turing Building, Oxford Road, University of Manchester, M13 9PL, UK.* michael.garrett@manchester.ac.uk
*   Corresponding Author

## Abstract

This study examines the hypothesis that the rapid development of Artificial Intelligence (AI), culminating in the emergence of Artificial Superintelligence (ASI), could act as a "Great Filter" that is responsible for the scarcity of advanced technological civilisations in the universe. It is proposed that such a filter emerges before these civilisations can develop a stable, multiplanetary existence, suggesting the typical longevity *(L)* of a technical civilization is less than 200 years. Such estimates for *L*, when applied to optimistic versions of the Drake equation, are consistent with the null results obtained by recent SETI surveys, and other efforts to detect various technosignatures across the electromagnetic spectrum. Through the lens of SETI, we reflect on humanity's current technological trajectory – the modest projections for *L* suggested here, underscore the critical need to quickly establish regulatory frameworks for AI development on Earth and the advancement of a multiplanetary society to mitigate against such existential threats. The persistence of intelligent and conscious life in the universe could hinge on the timely and effective implementation of such international regulatory measures and technological endeavours.

## 1. Introduction

One of the most puzzling results obtained by astronomers over the last 60 years is the non-detection of potential extraterrestrial "technosignatures" in astronomical data [e.g. 1-9]. These technosignatures are expected as a consequence of the activities of advanced technical civilisations located in our own and other galaxies e.g. narrowband radio transmissions, laser pulses, transiting megastructures, and waste-heat emission [10-12]. This "Great Silence", a term introduced by Brin [13], presents something of a paradox when juxtaposed with other astronomical findings that imply the universe is hospitable to the emergence of intelligent life. As our telescopes and associated instrumentation continue to improve, this persistent silence becomes increasingly uncomfortable, questioning the nature of the universe and the role of human intelligence and consciousness within it.

Various explanations for the great silence, and solutions to the related Fermi paradox [14] have been proposed [15]. The concept of a "great filter" [16] is often employed – this is a universal barrier and insurmountable challenge that prevents the widespread emergence of intelligent life. Examples of possible great filters are numerous, ranging from the rarity of abiogenesis itself, to the limited longevity of a technical civilization.

Most recently, Artificial Intelligence (AI) has also been proposed as another potential great filter and explanation for the Fermi Paradox [17,18]. The term AI is used to describe a human-made tool that emulates the "cognitive" abilities of the natural intelligence of human minds [18]. Recent breakthroughs in machine learning, neural networks, and deep learning have enabled AI to learn, adapt, and perform tasks once deemed exclusive to human cognition [19]. As AI rapidly integrates itself into our daily lives, it is reshaping how we interact and work with each other, how we interact with technology and how we perceive the world. It is altering communication patterns and personal experiences. Many other aspects of human society are being impacted, especially in areas such as commerce, health care, autonomous vehicles, financial forecasting, scientific research, technical R&D, design, education, industry, policing, national security and defence [20]. Indeed, it is difficult to think of an area of human pursuit that is still untouched by the rise of AI.

Many regard the development of AI as one of the most transformative technological developments in human history. In his BBC Reith Lecture (2021), Stuart Russell claimed that "the eventual emergence of general-purpose artificial intelligence [will be] the biggest event in human history [21]. Not surprisingly, the AI revolution has also raised serious concerns over societal issues such as workforce displacement, biases in algorithms, discrimination, transparency, social upheaval, accountability, data privacy, and ethical decision making [22-24]. There are also concerns about AIs increasing carbon footprint and its environmental impact [25].

In 2014, Stephen Hawking warned that the development of AI could spell the end of humankind. His argument was that once humans develop AI, it could evolve independently, redesigning itself at an ever-increasing rate [26]. Most recently, the implications of autonomous AI decision-making, have led to calls for a moratorium on the development of AI until a responsible form of control and regulation can be introduced [27].

Concerns about Artificial Superintelligence (ASI) eventually going rogue is considered a major issue - combatting this possibility over the next few years is a growing research pursuit for leaders in the field [28]. Governments are also trying to navigate a difficult line between the economic benefits of AI and the potential societal risks [29-32]. At the same time, they also understand that the rapid incorporation of AI can give a competitive advantage over other countries/regions – this could favour the early-adoption of innovative AI technologies above safeguarding against the potential risks that they represent. This is especially the case in areas such as national security and defence [33] where responsible and ethical development should be paramount.

In this paper, I consider the relation between the rapid emergence of AI and its potential role in explaining the "great silence". We start with the assumption that other advanced technical civilisations arise in the Milky Way, and that AI and later ASI emerge as a natural development in their early technical evolution. Section 2 addresses the threat posed by AI and section 3 considers how AI will progress in comparison to less well-developed mitigating strategies, in particular the development of a multiplanetary capability. Section 4 focuses on the short communicating lifetimes implied for technical civilisations and how this compares with the findings from SETI surveys. Section 5 advocates for the rapid regulation of AI and section 6 presents the main conclusions of the paper.

## 2. The threat posed by AI to all technical civilisations

AI has made extraordinary strides over the last decade. The impressive progress has underlined the fact that the timescales for technological advance in AI are extremely short compared to the timescales of Darwinian evolution [34]. AI's potential to revolutionize industries, solve complex problems, and simulate intelligence comparable to or surpassing human capabilities has propelled us into an era of unprecedented technological change. Very rapidly, human society has been thrust into uncharted territory. While the convergence of AI with other new technologies, including the Internet of Things (IoT) and robotics is already fuelling levels of apprehension about the future, also in terms of security issues [35] .

As noted by Yuval Harari, nothing in history has prepared us for the impact of introducing non-conscious super intelligent entities on the planet [36]. It is entirely reasonable to consider that this applies to all other biological civilisations located elsewhere in the universe. Even before AI becomes superintelligent and potentially autonomous, it is likely to be weaponized by competing groups within biological civilisations seeking to outdo one another [37]. The rapidity of AI's decision-making processes could escalate conflicts in ways that far surpass the original intentions. At this stage of AI development, it's possible that the wide-spread integration of AI in autonomous weapon systems and real-time defence decision making processes could lead to a calamitous incident such as global thermonuclear war [38], precipitating the demise of both artificial and biological technical civilisations.

While AI may require the support of biological civilisations to exist, it's hard to imagine that this condition also applies to ASI. Upon reaching a technological singularity [39], ASI systems will quickly surpass biological intelligence and evolve at a pace that completely outstrips traditional oversight mechanisms, leading to unforeseen and unintended consequences that are unlikely to be aligned with biological interests or ethics. The practicality of sustaining biological entities, with their extensive resource needs such as energy and space, may not appeal to an ASI focused on computational efficiency—potentially viewing them as a nuisance rather than beneficial. An ASI, could swiftly eliminate its parent biological civilisation in various ways [40], for instance, engineering and releasing a highly infectious and fatal virus into the environment.

Up to this point, we have considered AI and biological organisms as distinct from one another. Yet, on-going developments suggests that hybrid systems, may not be that far off. The question arises whether such advances could make biological entities more relevant to AI, perhaps preserving their existence into the future. This prospect seems unlikely. Brain-computer interfaces (BCIs) [41] may appear beneficial for enhancing biological organisms, but it's hard to see what long-term advantages AI would perceive in merging into a hybrid form. Indeed, there are many disadvantages including the complex maintenance requirements of biological systems, their limited processing capabilities, rapid physical decline, and vulnerability in harsh environments.

## 3. Multiplanetary mitigating strategies and technology progression

In our analysis thus far, we have assumed that AI and biological systems are co-located in the same limited physical space. As soon as this is no longer true, the existential threats that we have described in section 2 are no longer so stringent. For example, a multiplanetary biological species [42] could take advantage of independent experiences on different planets, diversifying their survival strategies and possibly avoiding the single-point failure that a planetary-bound civilisation faces.

A multiplanetary civilization could distribute its risk across several widely separated celestial bodies, reducing the likelihood of simultaneous destruction across all platforms. This distributed model of existence increases the resilience of a biological civilization to AI-induced catastrophes by creating redundancy. If one planet or outpost in space falls to a misalignment of AI's goals with biological interests, others may survive and immediately learn from these failures. Moreover, the expansion into multiple widely separated locations provides a broader scope for experimenting with AI. It allows for isolated environments where the effects of advanced AI can be studied without the immediate risk of global annihilation. Different planets or outposts in space could serve as test beds for various stages of AI development, under controlled conditions.

We know from our own experience that AI is progressing at a breath-taking pace. However, the same is not true of our own efforts to become a multi-planetary civilisation. Space is hard, and unfortunately, we seem to be a lot closer to achieving a technical singularity than realizing a truly multi-planetary, space faring capability.

The disparity between the rapid advancement of AI and the slower progress in space technology is stark. The technical singularity—a hypothetical point where AI surpasses human intelligence and capability—could occur within just a few decades, according to some predictions. In contrast, establishing a self-sustaining, multi-planetary human civilisation seems like a monumental task that may take many decades [43], possibly centuries. The essence of the problem lies in the nature of the challenges each domain faces. AI development is largely a computational and informational challenge, accelerated by the exponential growth of data and processing power. While AI can theoretically improve its own capabilities almost without physical constraints, space travel must contend with energy limitations, material science boundaries, and the harsh realities of the space environment. In addition, there are unsolved issues with respect to multiplanetary governance, biomedical and behaviour and logistical issues [44, 45, 46].

Given this potential universality of technological evolution, the implications for biological civilizations are profound. The race to develop AI could inadvertently prioritize advancements that lead to existential risks, overshadowing the slower-paced, yet arguably more vital, endeavour of becoming a multiplanetary species. Ironically, AI is likely to be a key tool in achieving the technical breakthroughs necessary to realise this goal.

## 4. Timescales and confrontation with the data

The scenario developed in section 3 suggests that almost all technical civilisations collapse on timescales set by their wide-spread adoption of AI. If AI-induced calamities need to occur before any civilisation achieves a multiplanetary capability, the longevity (L) of a communicating civilization as estimated by the Drake Equation [47], suggests a value of $L \sim 100$-200 years.

Let us consider the Drake Equation in more detail with a particular focus on the number *(N)* of radio-communicating technical civilizations in the galaxy:

$$N = R* \cdot fp \cdot ne \cdot fl \cdot fi \cdot ft \cdot L \qquad (1)$$

where $R*$ is the rate of star formation averaged over the lifetime of the Galaxy, *fp* is the fraction of stars with planetary systems, ne is the mean number of planets in each planetary system with environments favourable for life, *fl* is the fraction of such favourable planets on which life does in fact develop, *fi* is the fraction of such inhabited planets on which an intelligent civilisation arises, *ft* is the fraction of planets populated by an advanced technical civilisation and *L* is the lifetime of a radio communicating technical civilisation.

The first three astronomical terms of the equation are relatively well established ($R* \cdot fp \cdot ne \sim 0.1$ [48]) but the next three terms are not *(fl · fi · fc)*. Astronomers often assume highly optimistic values for these terms *( fl · fi · fc ~ 0.1)* while biologists suggest values many orders of magnitude smaller [49]. Even if we adopt the optimistic values, we derive for *N*:

$$N \sim 0.01\ L \qquad (2)$$

For values of *L ~ 100-200* years, we find *N ~ 1-2.*

A short communicative phase is therefore consistent with the null results from current SETI surveys. The window during which a technical civilisation can engage in detectable interstellar radio transmissions is extremely limited. In addition, if we assume the radio leakage emitted by emerging technical civilisations is like our own [50], the detection of other civilisations will be extremely challenging even for those located within our local stellar neighbourhood. The detection of more powerful directed signals (e.g. military radar) is of course possible across interstellar distances [51]. However, if only a handful of technical civilizations exist in the Milky Way at any given time, the probability of a detection occurring at cm-wavelengths within the very limited field of view offered by the current generation of large single-dishes and beam-formed arrays is minimal. An "all-sky" capability or something approaching this would be required, and this surpasses the capabilities of current SETI radio instruments by a substantial margin. SETI researchers may need to consider the types of instruments required to conduct meaningful surveys – field of view is a metric which is often overlooked compared to raw sensitivity and total bandwidth. [52].

We also note that a post-biological technical civilisation would be especially well-adapted to space exploration [53, 54], with the potential to spread its presence throughout the Galaxy, even if the travel times are long and the interstellar environment harsh. Indeed, many predict that if we were to encounter extraterrestrial intelligence it would likely be in machine form [55]. Contemporary initiatives like the Breakthrough Starshot programme [56] are exploring technologies that would propel light-weight electronic systems toward the nearest star, Proxima Centauri. It's conceivable that the first successful attempts to do this might be realised before the century's close, and AI components could form an integral part of these miniature payloads. The absence of detectable signs of civilisations spanning stellar systems and entire galaxies (Kardashev Type II and Type III civilisations) further implies that such entities are either exceedingly rare or non-existent [8,9], reinforcing the notion of a "Great Filter" that halts the progress of a technical civilization within a few centuries of its emergence.

## 5. AI regulation

The field of SETI aims not only to search for intelligent life beyond Earth but also holds up a mirror to humanity, encouraging us to reflect on our own technological progression and potential futures. By examining the possibilities of alien civilisations, SETI helps us contemplate the long-term sustainability of our own civilisation, the potential risks we face, and how we might navigate and overcome future challenges.

Presently, the AI we currently encounter in every-day life largely operates within human-established constraints and objectives. Nevertheless, progress is being made in creating systems that can augment and optimize various facets of their own development [57]. The next stage will see AI systems independently innovate and refine their own design without human intervention. The potential for AI to operate autonomously raises many ethical and moral quandaries but it is surely only a matter of time before this occurs. Tests are already being conducted in military settings, and the proliferation of Lethal Autonomous Weapons Systems (LAWS) by rogue nations or covert organisations is surely inevitable [58]. We stand on the brink of exponential growth in AI's evolution and its societal repercussions and implications. This pivotal shift is something that all biologically-based technical civilisations will encounter. Given that the pace of technological change is unparalleled in the history of science, it is probable that all technical civilisations will significantly miscalculate the profound effects that this shift will engender [21,26,36].

There can be little doubt that AI and in particular ASI present a massive challenge to the longevity of our technical civilisation and likely all technical civilisations that arise in the cosmos. This naturally leads us to the thorny matter of AI regulation and control. While industry stakeholders, policymakers, individual experts, and their governments already warn that regulation is necessary [27], establishing a regulatory framework that can be globally acceptable is going to be challenging. In the meantime, AI continues to progress. In particular, nations have diverse cultural, economic, and societal priorities, leading to varied perspectives on the governance of AI [59]. Geopolitical interests cannot be ignored – even if comprehensive regulations were adopted, some nations will be tempted to bend the rules. In addition, rapid advances in AI will likely outpace any agreed regulatory frameworks, raising concerns that the latter will always lag well behind new and unanticipated advances in the field.

Ensuring compliance and accountability in AI development and deployment also poses significant challenges. The decentralised nature of AI development, the enormous size of the global AI research community spread across almost every research domain will further complicate the oversight and enforcement of regulations. In short, regulation of this new technology is going to be very difficult, if not impossible to achieve. Without practical regulation, there is every reason to believe that AI could represent a major threat to the future course of not only our technical civilisation but *all* technical civilisations.

## 6. Conclusions

The rapid development of AI presents a formidable challenge to the survival and longevity of advanced technical civilisations, not only on Earth but potentially throughout the cosmos. The pace at which AI is advancing is without historical parallel, and there is a real possibility that AI could achieve a level of superintelligence within a few decades. The development of ASI is likely to happen well before humankind manages to establish a resilient and enduring multiplanetary presence in our solar system. This disparity in the rate of progress between these two technological frontiers is a pattern that we can expect to be repeated across all emerging technical civilizations.

This raises questions about the inevitability of civilisations unwittingly triggering calamitous events that lead to the demise of both a biological and post-biological technical civilisation. The potential of ASI to serve as a "Great Filter" compels us to consider its role in the broader context of our civilization's future and its implications for life throughout the galaxy. If ASI limits the communicative lifespan of advanced civilizations to a few hundred years, then only a handful of communicating civilisations are likely to be concurrently present in the Milky Way. This is not inconsistent with the null results obtained from current SETI surveys and other efforts to detect technosignatures across the electromagnetic spectrum.

If SETI also serves as a lens through which we can examine our own technological trajectory and societal challenges, the urgency of establishing comprehensive global AI regulations cannot be overstated. It behoves us to engage with these issues proactively, to develop and enforce prudent regulatory measures, and to strive for a balance between harnessing the benefits of AI and safeguarding against the existential risks it may pose. As we stand on the precipice of a new era in technological evolution, the actions we take now will determine the trajectory of our civilization for decades to come. The implied longevity timescales for the scenarios described here (approximately 100-200 years), underscores the necessity for our own technical civilization to intensify efforts to control and regulate AI. The continued presence of consciousness in the universe may depend on the success of strict global regulatory measures.

### Declaration of competing interest

The author declares that he has no known competing financial interests or personal relationships that could have appeared to influence the work reported in this paper.

### Acknowledgements

I'd like to thank the referee for very useful comments on the original submission. This research did not receive any specific grant from funding agencies in the public, commercial, or not-for-profit sectors.